# Multikast rutiranje open-source platformom - XORP

Petar Bojović, Katarina Savić, Aleksandra Smiljanić

*Sadržaj* — Integracijom softverskog rutera u embedded sisteme dobija se mogućnost najsavremenijih rutera, po znatno pristupačnijoj ceni. Servisi prenosa TV i radio signala preko IP mreže, zaživljavaju tek korišćenjem *multikast*[1] protokola za rutiranje. *Multikast* rutiranje[2] je trenutno funkcija samo skupih hardverskih rešenja. XORP open-source platforma nudi *multikast* rutiranje kroz softverski ruter, sa mogućnošću integracije u jeftine embedded platforme[3].

*Ključne reči* — **Alix, Soekris, embedded, multikast, rutiranje, XORP, VLC,**

## I. Uvod

Svedoci smo sve većeg razvoja softverskih platformi za rutiranje. Osnovna inspiracija programerima za softverske alternative hardverskim rešenjima je upravo cena kako razvoja tako i proizvodnje proizvoda. Jednom razvijem softverski projekat može se koristiti u različitim komercijalnim proizvodima. Danas smo svedoci toga da mnogi proizvodi koji su do skoro postojali samo u hardverskoj varijanti, postoje sad, više ili manje popularni, i u softverskoj varijanti. Niža maloprodajna cena uređaja koji glavnu funkcionalnost obavlja programski, je glavni razlog zbog kojeg će softverske varijante imati svoje zasluženo mesto na tržištu.

Međutim, i softverskoj varijanti je potreban hardver koji će izvršavati zadatu funkcionalnost. Taj hardver je najčešće neki opšti tip arhitekture računarskog sistema, koji se sastoji od centralnog procesora (CPU), radne memorije (RAM), sistema za skladištenje podataka (storage card), i periferija. Ta hardverska arhitektura je zadužena za sve operacije koje obezbeđuje softver. Lako je onda zaključiti da su maksimalne performanse ograničene

P. Bojović, Računarski fakultet, Beograd, Srbija (telefon: 381-11-2627613; faks: 381-11-2623287; e-mail: petar.bojovic.paxy@gmail.com).
K. Savić, Elektrotehnički fakultet, Univerzitet u Beogradu (catrins@gmail.com)
A. Smiljanić, Elektrotehnički fakultet, Univerzitet u Beogradu (aleksandra@etf.rs).

mogućnošću te arhitekture. Svaka softverska naredba sastoji se od niza operacija koje CPU treba da izvrši. Kod hardverske implementacije funkcionalnih jedinica, dobro je poznato da performanse tih jedinice određuje njihova arhitektura. Kod softverske implementacije performanse su zajedničke za celu platformu. Ipak, prednost softverske implementacije je da se razvoj nekog projekta mnogo lakše i efikasnije vrši. Dogradnja, korekcija i testiranje je neuporedivo lakše, jeftinije, i dostupnije kod softverskih projekata.

Kada se radi o uređajima za rutiranje mrežnog saobraćaja, postoji mnogo varijanti i softverskih i hardverskih rutera. Hardverska rešenja rutera su poznata po izuzetno dobrim performansama ali i veoma visokoj ceni. Softverske alternative su u poslednje vreme, zbog svoje cene koja je i do 1000 puta manja, postale popularne svuda, a posebno u siromašnijim okruženjima. Iako po performansama se ne mogu meriti sa hardverskim, softverski sistemi su ih brojčano nadmašili. Kako bi se našao što bolji odnos cene i performansi, teži se uspostavljanju što boljeg balansa između hardverskih i softverskih funkcionalnih jedinica. Stoga najbolje rešenje je upravo hibridni sistem. One funkcionalne jedinice koje najviše utiču na ukupne performanse uređaja neophodno je implementirati hardverski. Napredne funkcionalnosti ruter sistema treba implementirati softverski jer su u stalnom razvoju i ne utiču mnogo na performanse.

XORP (eXtensible Open-source Routing Platform) je jedan od projekata softverske implementacije ruter sistema. Namenjen je za korišćenje na računarskim arhitekturama koje mogu da pokrenu neki od popularnih Operativnih sistema. U ovom radu biće prikazano *multikast* rutiranju što predstavlja jednu od naprednih mogućnosti XORP softverske platforme.

## II. HARDVERSKE EMBEDDED PLATFORME

Testiranje XORP softverske platforme za rutiranje se može izvršiti na bilo kojoj arhitekturi za koji postoje popularni operativni sistemi tipa Windows, Linux ili BSD. Za potrebe simulacije može poslužiti i standardni PC računar. U svrhu testiranja funkcionalnosti XORP platforme, izvršena je priprema i instalacija XORP-a na PC računaru preko virtualnih mašina, ali druge dve embedded platforme. Embedded platforme su hardverski sistemi različitih arhitektura koje ne poseduju sve funkcionalnost kao PC arhitektura. Najčešće ne poseduju grafičku kartu, niti periferije tipa hard disk i CD-ROM.

### A. *Soekris Net4801*

CF (Compact Flash)[4] kartica se koristi kao medijum za skladištenje podataka, tako da se operativni sistem i aplikacije nalaze na njoj. Zbog toga

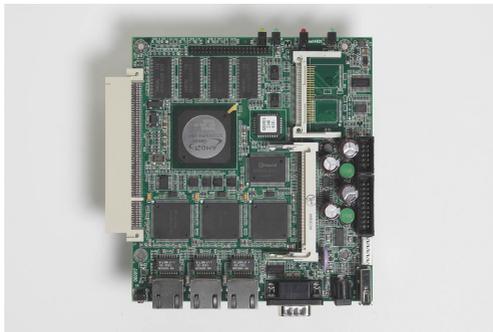

*Slika III.1 Soekris Net4801 platforma*

| | |
|---|---|
| CPU: 233 MHz NSC SC1100 single chip processor | USB: 2 x USB 1.1 |
| RAM: 128 MB SDRAM integrated | COM: 1 x DB9, 1 x 10 pin header |
| BIOS: 4Mbit BOOT | LED: Power, Activity, Error |
| Storage: CompactFlash I/II | PCI: 1 x PCI slot 3.3 V, 1 x MiniPCI slot type III |
| LAN: 3 x 10/100 FastEthernet RJ45 | Power: 6-20V max 15 Watts |

je neophodna za rad sistema. Korišćena je kartica od 1 Gb. Za proširenu funkcionalnost rutera nabavljena je dodatna kartica koja pored 3 priključka za LAN omogućava priključak još 4 mrežna interfejsa. Ovo je omogućeno PCI karticom sa 4 FastEthernet porta.

Što se tiče operativnog sistema bitno je napomenuti da procesor koji se nalazi na ovom sistemu podržava i386 skup instrukcija. To znači da ovaj sistem može da podrži sve operativne sisteme koji imaju podršku za pomenuti skup instrukcija tj. arhitekturu.

*B. PC Engines, Alix 2d2*

Ovaj sistem takođe koristi CF karticu kao kao medijum za skladištenje podataka. Za potrebe instalacije operativnog sistema i drugog softvera korišćena je kartica od 2 GB. Alix 2c2 poseduje Geode procesor koji koristi i386 skup instrukcija tako da podržava sve operativne sisteme namenjene ovoj arhitekturi.

### III. SOFTVER

Kada se radi o softverskoj varijanti funkcionalnih uređaja postoji nekoliko softverskih nivoa koje je neophodno implementirati (Slika IV.1). Sam hardver pruža samo računarsku arhitekturu i mogućnost korišćenje te arhitekture preko asemblerskih instrukcija koje dati procesor razume. Ako bismo hteli da pišemo softver tako da se direktno bazira na hardverskoj arhitekturi, morali bismo da prilagodimo instrukcije za taj specifičan hardver, koristeći tačan format instrukcija koji taj hardver zahteva. Tako napisan program ne bi bio

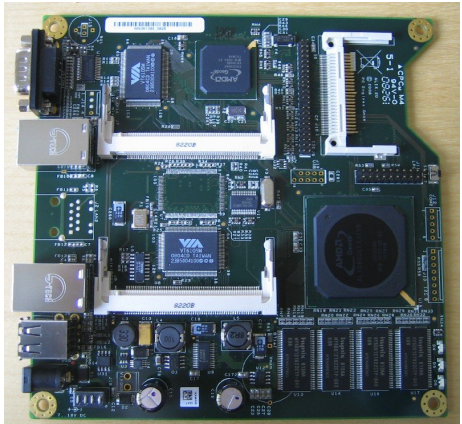

*Slika III.2 Alix 2d2 platforma*

CPU: 500 MHz AMD Geode LX800  
RAM: 256 MB DDR DRAM  
Storage: CompactFlash socket  
LAN: 2 x FastEthernet 10/100 Mbps  
USB: 2 x USB 1.1  
COM: DB9 serial port  
LED: 3 x panel LED  
PCI: 2 x MiniPCI slot, LPC bus  
Power: 7-20 V, DC jack

portabilan, tj. ne može se lako prilagoditi drugoj hardverskoj arhitekturi. Česta je pojava da zbog tehnološkog napretka prestane da se proizvodi određeni hardverski deo, pa je proizvođač primoran da nađe alternativu. Bilo bi jako nezgodno da zbog toga mora da se vrši prilagođavanje softvera.

Kako bi se izbegla tolika zavisnost softvera od hardvera na kome se izvršava, neophodno je koristiti još jedan softverski nivo koji bi zanemario hardver za dalji razvoj softvera. Operativni sistem je nivo koji nam pruža određenu nezavisnost našeg softvera od „ležećeg" (underlying) hardvera. Operativni sistemi bazirani na Linux i BSD kernelu su uglavnom open-source sistemi, što znači da se slobodno može preuzeti izvorni kod, kompajlirati i instalirati sistem.

Podrška operativnog sistema različitim embedded platformama se vrši prilagođenjem koda toj arhitekturi i kompajliranjem. Za najčešću arhitekturu računara, i386, postoje već kompajlirani operativni sistemi u bilo kojoj distribuciji.

Pored operativnog sistema, softverski paketi često vrše svoju funkcionalnost koristeći neki drugi softverski paket. Ova relacija se naziva zavisnost (dependences). Prema pomenutoj softverskoj hijerarhiji nakon instalacije operativnog sistema, sledeći korak je pristupanje kompajliranju i instalaciji XORP platforme. XORP poseduje određene zavisnosti. Da bi se kompajlirao XORP traži da se instalira OpenSSL paket.

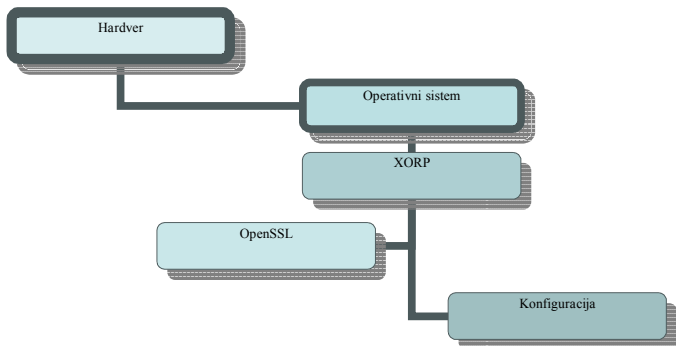

*Slika IV.1 Dijagram organizacije softvera*

Nakon instalacija svih zavisnosti, može se pristupiti kompajliranju i instalaciji XORP-a. Poslednji korak predstavlja konfigurisanje XORP-a za potrebnu funkcionalnost.

## IV. OPERATIVNI SISTEM

Internet zajednica XORP (XORP *Community*) poseduje brojne izveštaje ljudi koji sa uspehom koriste XORP na različitim operativnim sistemima. Prijavljeno je da XORP radi na sistemima: FreeBSD, OpenBSD, DragonFlyBSD, NetBSD, Linux, MacOS X, Windows Server 2003. Ipak, treba biti pažljiv prilikom izbora operativnog sistema, jer kod svake varijante se mogu dogoditi potpuno drugačiji problemi u procesu instalacije.

XORP platforma poseduje *Live* CD distribuciju koja je namenjena za potrebe testiranja platforme na PC računarima. To je rešenje u kom je na jednom CD disku smešten operativni sistem, svi paketi od kojih XORP zavisi, sam XORP, neka *default* konfiguracija, kao i mehanizmi za *save* i *load* konfiguracije na USB fleš disk. To je idealan paket, za one koji žele XORP samo da isprobaju na svom kućnom računaru, a pri tom ne poseduju neophodna znanja za rukovanje Linux ili BSD sistemima. Da bi se koristio *Live* CD XORP-a, neophodno je prvo preuzeti sliku *Live* CD sa XORP sajta (http://www.xorp.org/livecd.html), zatim narezati CD prema slici. Podesiti računar da podiže sistem sa CD uređaja i ubaciti bilo kakav USB fleš disk. Sa CD-a će se pokrenuti procedura za podizanje operativnog sistema. Pošto je kompletan sistem i konfiguracija smeštena na CD-u koji je *read only*, USB fleš se koristi za čuvanje naše konfiguracije (koju će učitati prilikom sledećeg restarta računara). Kada se podigne operativni sistem, podiže se i XORP automatski. Da bi konfigurisali XORP da radi ono što želimo potrebno je aktivirati konzolu XORP-a sa komandom *xorpsh*.

Ukoliko ne želimo da koristimo XORP samo za svrhe testiranja, mora se izvršiti instalaciju na nekom od operativnih sistema. Tada postupak opisan upotrebom *Live* CD-a nije posebno koristan.

Pošto želimo da osposobimo dve *embedded* platforme da nam budu softverski ruteri, neophodno je podići operativni sistem na oba uređaja. Radi demonstracije uzeti su različiti operativni sistemi za ove dve *embedded* platforme.

A. *Alix 2c2 – Voyage Linux 0.6.0*

Voyage je operativni sistem baziran na Linux kernelu i predstavlja Debian distribuciju koja se prilagođena za *embedded* platforme. Prilikom izbora operativnog sistema na *embedded* platformama, koje imaju znatno slabije resurse nego PC računari, bitno je izabrati sistem koji neće previše za svoje potrebe koristiti raspoložive resurse. Baš zbog tog razloga postoji operativni sistem Voyage namenjen Soekris i Alix (nekada poznat kao Wrap) pločama.

Treba istaći da je bilo koji Linux, pa i Voyage, ipak sistem koji više konzumira resurse nego bilo koji BSD. Međutim Alix ploča poseduje znatno jači procesor nego Soekris, pa korišćenje Linux-a može doprineti bolje iskorišćenje resursa nego što bi to bio slučaj da se koristi BSD.

Već je pomenuto da *embedded* platforme ne poseduju grafičku kartu, tj. grafički interfejs ka korisniku. Zbog toga pristup samom uređaju se ne može vršiti na način na koji se to radi preko PC računara (monitor, tastatura i miš).

Pristup i kontrola uređaja se vrši preko serijskih COM portova. Potrebno je povezati uređaj DB9 (devetopinskim serijskim) kablom sa PC računarom. Noviji PC računari ne poseduju više standardan DB9 port već je neophodno nabaviti adapter sa DB9 na USB port.

Komunikacija sa uređajem, nakon fizičkog povezivanja, se uspostavlja povezivanjem na odgovarajući COM port (često COM1, proveriti port u Device Manager) preko terminalnih aplikacija kao što su HyperTerminal, Putty i sl.

Prilikom povezivanja na COM port neophodno je definisati baud rate[4] koji zavisi od uređaja do uređaja. Najčešće se prvo proba sa brzinom od 9600 b/s, pa 19200 b/s, pa 38400 b/s. Ukoliko se nakon povezivanja dobiju ne razumljivi karakteri znači da je promašen baud rate.

Kada uređaj dobije napajanje, prvih nekoliko sekundi izvršava se BIOS uređaja. Najčešće odmah nakon paljenja BIOS vrši test ispravnosti resursa prikazujući rezultat na COM portu. Najbolji test da li je uspešno pristupljeno uređaju jeste povezivanje na odgovarajući COM port sa brzinom od 38400 b/s i restartovanje uređaja prekidajući napajanje. Nakon toga treba da se dobije odziv BIOS-a koji testira integrisani hardver.

Ono što je još neophodno primetiti kod *embedded* platformi je to što one ne poseduju periferije kao što su CD ili DVD ROM. Zbog toga je procedura instalacije operativnog sistema potpuno drugačija.

Sistem se može instalirati na jedan od dva načina:
1. Kopiranjem fajlova već instaliranog sistema na CF karticu
2. Instalacijom preko mreže – PXE (Preboot eXecution Environment)[1][4]

Dupliciranje već instaliranog sistema nikad nije preporučljivo. Postoje mnoge stvari unutar operativnog sistema koje moraju da budu unikatne samo za taj sistem. Praksa je pokazala da ukoliko se kopira ceo operativni sistem i promeni se makar minimalna stvar u hardverskoj postavi mogu se desiti razni teško rešivi problemi.

Instalacija preko PXE metode predstavlja gotovo identičan način instalacije kao i kod klasičnog PC računara upotrebom periferija, samo što je izvor mrežni resurs. Instalacija preko PXE zahteva:
1. DHCP server[2]
2. TFTP (*Trivial* FTP) server[2]
3. PXE sposoban klijent – PXE mrežnu karticu

Pored PXE metode za instalaciju potreban je još FTP ili HTTP server koji bi pružio fajlove potrebne za instalaciju preko mreže. Svi operativni sistemi ne podržavaju PXE proceduru instalacije. Oni koji podržavaju PXE instalaciju, imaju drugačiju proceduru.

Ukoliko se želi da instalira Voyage Linux na Alix ploču preko PXE metode procedura je sledeća:
- Preuzeti sliku Voyage Live CD (http://linux.voyage.hk/?q=download)
- Narezati sliku sistema na CD
- Pokrenuti PC računar sa *Live* CD-a Voyage Linux-a
- Kada se završi učitavanje sistema pokrenuti skriptu koja je prilagođena za instalaciju Voyage na *embedded* sistemima preko PXE: *voyage-pxe start 38400*

Skripta podiže DHCP server pružajući unapred definisanu IP adresu, kao i parametre za PXE BOOT fajl i TFTP server. Preko TFTP servera pruža fajlove koji su namenjeni za pokretanje instalacije sistema.

Dalja instalacija se vrši na Alix ploči:
- Izvrši se povezivanje (drugim računarom) na COM port preko kog je povezana ploča
- Poveže se LAN1 port Alix ploče sa mrežnom kartom računara gde je podignut Voyage sa PXE servisom. (možda je neophodan cross-over mrežni kabl, proveriti da li sijaju lampice od mreže)
- Restartuje se Alix uređaj i dok je još u BIOS fazi pritisnuti dugme

DELETE
- U konfiguraciji BIOS-a podesiti da prvi BOOT uređaj bude LAN
- Nakon izlaska iz BIOS-a pokreće se PXE klijent na mrežnoj kartici
- Pronalazi se DHCP server i uzima IP adresu, TFTP IP adresu servera, i ime fajla za podizanje instalacije
- Kad se prevuku potrebni fajlovi za pokretanje instalacije, pokreće se instalaciona procedura
- Instalacija operativnog sistema se obavlja automatski
- Sistem pita za nekoliko ključnih podataka, kao što su naziv uređaja, glavna šifra, i sl.
- Nakon instalacije sistema, tj. prvog restarta treba ponovo ući u BIOS i isključiti podizanje sa mreže.

Operativni sistem Voyage Linux je instaliran na CF karticu. U BIOS-u je podešeno da se automatski pokreće sistem sa CF kartice. Voyage Linux instalira GRUB kao *boot loader*, tako da je prozor izbora operativnog sistema prvo na šta se nailazi kad počne pokretanje sa CF kartice. Nakon uspešnog podizanja sistema, potrebno je prijaviri se sa nalogom *root* i šifrom koju ste postavili u toku instalacije.

Procedura za instalaciju Voyage Linux je ovime završena. Ostalo je još konfigurisati rad sistema sa mrežnim parametrima. Ukoliko je aktivan DHCP servis, sistem će prilikom podizanja zahtevati IP adresu. Pregled dodeljene IP adrese možete izvršiti preko komande: *ifconfig*. Postavljanje IP adrese ručno se vrši na sledeći način:

*ifconfig eth0 192.168.1.2 netmask 255.255.255.0*

gde je eth0 – logički naziv mrežnog interfjesa u Linux-u.

Ukoliko se želi da se interfejs automatski konfiguriše na neku statičku IP adresu nakon svakog podizanja sistema, potrebno je editovati fajl */etc/networks/interfaces.*

### B. Soekris net4801 – OpenBSD 4.4

XORP platforma je razvijana na operativnom sistemu BSD. Kernel operativnog sistema BSD je znatno manji i jednostavniji nego kernel operativnog sistema Linux. To omogućava da sam sistem ne zauzima mnogo resursa prilikom izvršavanja, ali i umanjuje korisnost za neke napredne funkcije. Kako Soekris ploča sa kojom radimo, ima lošiji procesor nego Alix ploča, korišćen je BSD sistem na ovoj arhitekturi.

Kao i Alix ploča, tako i Soekris ne poseduje grafičku kartu, niti drugi ulazni interfejs kao što je tastatura. Zbog toga, način pristupa uređaju je identičan kao i kod Alix ploče, tj. pristupa se preko konzolnog porta. Za pristup Soekris ploči neophodno je povezati serijski COM port uređaja sa PC računarom (preko *serial* na USB *converter*). Soekris pločama se pristupa preko terminalnih programa kao što je Putty, konektujući se na odgovarajući COM port i brzinom od 9600 b/s.

Soekris ploča takođe poseduje POST BIOS test ispravnosti hardvera, koji se aktivira prvih nekoliko sekundi kada uređaj dobije napajanje. Kada se želi da se isproba povezivost uređaja sa PC računarom, on se poveže na odgovarajući COM port i priključi mu se uređaj na napajanje.

Procedura instalacije operativnog sistema je konceptualno ista kao i kod Alix ploče, razlika je u tome što OpenBSD nema automatizovane skripte za pokretanje neophodnih servisa kako bi PXE funkcionisao, već se servisi moraju ručno podesiti u na nekom drugom Linux ili BSD sistemu.

## V. PREDUSLOVI ZA KOMPAJLIRANJE XORP PLATFORME

Prvi korak ka kompajliranju XORP platforme je preuzimanje izvornog koda. Poslednja verzija izvornog koda se može preuzeti sa matičnog sajta na adresi: http://www.xorp.org/downloads.html. Verzija sa kojom je izvršeno testiranje je 1.6. Na sajtu se mogu pronaći fajlovi sa izvornim kodom ali i instalaciona verzija za Windows. Takođe stoji upozorenje da za Windows i MacOS nije napravljena potpuna podrška, pa da *multikast* rutiranje neće raditi. Sama datoteka u kojoj se nalazi izvorni kod je veličine 7.5 MB u .tar.gz arhivi.

Međutim, iako je datoteka relativno mala, XORP zahteva mnogo veći prostor na medijumu za skladištenje podataka da bi se uspešno obavilo kompajliranje. Na matičnom sajtu XORP-a piše da je neophodno obezbediti oko 1.4 GB prostora da bi se otpakovao i kompajlirao program.

Tu se dolazi do problema. Embedded platforme imaju relativno malu memoriju za skladištenje podataka. CF kartica je jedna od skupljih fleš memorija jer ima znatno brži odziv od drugih, i znatno veći broj upisa po ćeliji (duži životni vek). Na Alix ploči se nalazi kartica od 2 GB, a na Soekris, 1 GB. U oba slučaja nakon instalacije svih potrebnih zavisnosti, neće ostati još 1.4 GB za kompajliranje izvornog koda XORP-a.

Kao rešenje iskorišćen je USB fleš memorija od 4 GB, čija je cena višestruko manja od cene CF kartice. Obe ploče poseduju USB port, što omogućava korišćenje USB fleš memorije. USB fleš memorija koja je prethodno korišćena u Windows-u tj. formatirana na FAT particiju može da se koristi i u Linux-u i BSD-u. Međutim, nije preporučljivo vršiti kompajliranje direktno na fleš memoriju koja je tako formatirana. Razlog je upravo to što su mehanizmi za kompajliranje prilagođeni Linux *file* sistemima, pa se koriste i opcije koje postoje samo u Linux *file* sistemu. Zbog toga je potrebno formatirati USB fleš na Linux particiju *ext2/ext3*.

Formatiranjem fleš memorije pripremljena je za korišćenje u svrhu kompajliranja na Linux/BSD sistemu. Ali, neophodno je i povezati (*mount*) memoriju sa operativnim sistemom. Kod Windows-a, kao i složenijih distribucija Linux-a, prilikom ubacivanja USB fleša automatski se vrši povezivanje, pa dobijamo posebno slovo u Windows-u, tj link u Linux-u. Kod Voyage linux-a i OpenBSD-a povezivanje sa USB fleš memorijom ne pokreće se automatski već je neophodno pozvati određenu komandu.

Prvi put kada se želi da poveže sistem sa USB fleš memorijom neophodno je kreirati direktorijum sa kojim ćete povezati memoriju. To treba uraditi preko komande: *mkdir /media/flash* – ako se želi da pristupi flešu kao direktorijumu /media/flash. Nakon toga povezivanje USB fleš memorije sa sistemom se vrši komandom:

*mount /dev/sda1 /media/flash* – u Linux Voyage
*mount –t ext2fs /dev/sda1i /media/flash* – u OpenBSD

Kod logičkog uređaja *sda* pristupa se *sda1* zbog toga što se pristupa prvoj particiji USB memorije. Ukoliko postoji više particija promenom broja može se pristupiti drugim partricijama. U OpenBSD sistemu koristi se *sda1i* kao pokazivač na particiju na *sda* drajvu.

Komandu za povezivanje fleša je neophodno uneti svaki put kad se ubaci USB fleš ili restartuje sistem. Posle toga može da se pristupi sadržaju USB memorije preko komande *cd /media/flash*.

USB fleš memorija koristi se samo privremeno kako bi se kompajlirala i izvršila instalacija XORP-a. Nakon toga svi potrebni fajlovi će se nalaziti na CF kartici. Gotovi kompajlirani fajlovi zauzimaju negde oko 40 MB, tako da nije problem smestiti ih na male CF kartice.
Izvorni kod XORP-a je neophodno smestiti u povezan direktorijum. To se može uraditi ili direktno preuzimanjem sa interneta na fleš preko komandi:

*cd /media/flash*
*wget http://www.xorp.org/releases/1.6/xorp-1.6.tar.gz*

ili koristeći WinSCP aplikaciju koja preko SSH prenosi datoteku na platformu. Ako koristite WinSCP, datoteka se prvo preuzme na računar, pa zatim smesti u /media/flash direktorijum.

Kada se datoteka sa izvornim kodom nalazi na USB fleš memoriji, potrebno ju je otpakovati komandom: *tar –xzf xorp-1.6.tar.gz* Otpakivanje može potrajati prilično dugo jer se koristi spora USB fleš memorija, a arhiva sadrži mnogo datoteka. Otpakovana verzija zauzima oko 1 GB prostora.

Na ovaj način je pripremljen izvorni kod XORP-a za kompajliranje. Ovime se završava standardni, predvidivi deo instalacionog procesa. Ono što je uvek nezgodno kod kompajliranja aplikacija u Linux-u jeste obezbediti sve potrebne zavisnosti (*dependeces*). U dosta slučajeva dešava se da isti postupak na dve mašine daje različite rezultate. Zbog toga ne postoji uputstvo koje se može tačno ispratiti i kao rezultat dati uspešno kompajliranje.

Prema uputstvu koje postoji na matičnom sajtu http://www.xorp.org/getting_started.html za kompajliranje i instalaciju neophodno je pokrenuti samo dve komande u XORP direktorijumu:

*./configure*
*gmake*

Ako se koristi neka od Linux distribucija koje dolaze u kompletnom paketu (nekoliko DVD-a) možda će kompajliranje proći po ovom uputstvu. U slučaju skraćenih verzija, sigurno ne.

Za pokretanje procedure razrešavanja zavisnosti neophodnih za kompajliranje XORP-a potrebno je krenuti sledećim redom:
- ući u direktorijum sa XORP kodom  - *cd /media/flash/xorp-1.6*
- pokrenuti *./configure*

Skripta *configure* služi da proveri okruženje, tj. operativni sistem, platformu i druge zavisnosti, kako bi konfigurisala program koji kompajlira.
Kada skripta naiđe na program, zavisnost, komponentu koja nedostaje, a bitna je za kompajliranje XROP-a, ona prekine izvršavanje. Tada je neophodno prvo rešiti taj problem, pa tek onda ponovo pokrenuti *configure* skriptu.

Ukoliko skripta prekine izvršavanje, poslednja linija objašnjava šta nedostaje. Ako se nema iskustva sa paketima u Linux/BSD sistemu, najbolje je prekopirati poruku sa greškom, i potražiti rešenje na Google. U gotovo 100% slučajeva na forumima se može naći da je neko imao sličan problem, te su iskusniji korisnici već dali savet za rešenje problema.

Jedan od problema za kompajliranje je ukoliko nije instaliran OpenSSL programski paket. Tada će se skripta zaustaviti u momentu testiranja postojanosti tog paketa. Većina dobro poznatih paketa može se instalirati direktno bez kompajliranja, jer postoji kompajlirana, stabilna, verzija u repozitorijumu – bazi kompajliranih programa.

Ukoliko se na Linux-u želi da instalira neki program iz repozitorijuma potrebno je izvršiti sledeće komande:
*apt-get update* – komanda kojom se preuzimaju podaci o svim kompajliranim verzijama iz centrale baze podataka *apt-get install openssl* – komanda kojom se zahteva da sistem preuzme i instalira već kompajliranu verziju OpenSSL-a.

Naravno, preduslov da *apt-get* radi, jeste da ima izlaz na internet, tj. pravilno podešene parametre IP adrese i *default* rutu. Ako se ne koristi DHCP, neophodno je dodati statičku *default* rutu sa komandom:
*route add default gw 192.168.1.254 eth0*
Linux već poseduje datoteku sa informacijom gde se na internetu nalazi repozitorijum kompajliranih programa.

Kod OpenBSD takođe postoji repozitorijum kompajliranih programa, ali sam sistem nema svoju bazu, već mu je neophodno podesiti putanju do baze svaki put pre korišćenja. Potrebno je podesiti promenljivu PKG_PATH tako da ukazuje na repozitorijum, i to se radi na sledeći način:
*export PKG_PATH=ftp://ftp.openbsd.org/pub/OpenBSD/4.4/packages/i386/*
Pomenuta putanja je za OpenBSD 4.4, i arhitekturu i386. Potrebno je proveriti te parametre pre instalacije softvera iz repozitorijuma, jer u suprotnom instalacija može da ošteti sistem.

Nakon podešavanja ove putanje instalacija paketa se vrši preko komadne:
*pkg_add openssl* – čime će sistem preuzeti kompajliranu datoteku i instalirati je.

Kod instalacije preko repozitorijuma, ukoliko željenom programu nedostaju neke zavisnosti (*dependences*), sistem automatski preuzima i te pakete, te ih instalira pre instalacije zahtevanog programa.

I kod OpenBSD-a je preduslov izlaz na internet, pa ako se ne koristi DHCP za dodelu *default* rute, ručno se dodaje preko komande:
*route add default 10.130.128.1*

Napomena: Ne treba pokušavati da se instalira XORP iz repozitorijuma, jer verzije koje se nalaze u repozitorijumu su oštećene, pa mogu da oštete sistem.

Proces razrešavanja svih zavisnosti koje traži XORP zavisi od stanja operativnog sistema koji se koristi.

Nabolji postupak je upravo opisan, tj. kada se prekine *configure* skript, potraži se rešenje tog problema preko npr. Google-a, i vidi se koji paket nedostaje. Kada se to sazna instalira se taj paket preko:
*apt-get install paket* ili *pkg_add paket* i ponovo pokrene *configure* skripta.

## VI. KERNEL PREDUSLOV – LINUX/BSD

Skripta *configure* neće prijaviti problem ukoliko kernel operativnog sistema nije podešen da podržava *multikast* i *multikast* rutiranje. Naime, XORP bez podrške u kernelu će raditi dokle god se ne koristi *multikast*.

OpenBSD 4.4 već dolazi sa podrškom za *multikast* rutiranje u kernelu, pa nije neophodno vršiti bilo kakve intervencije. Linux gotovo u svim distribucijama dolazi bez podrške za *multikast* rutiranje. Zato je neophodno da se izvrši rekonfiguracija i rekompajliranje kernela da bi se aktivirala podrška za *multikast*. Postupak podešavanja Linux kernel-a za korišćenje *multikast* rutiranja mora da se potraži u dokumentaciji aktuelnog kernel-a.

### VII. KOMPAJLIRANJE XORP PLATFORME

Na matičnom sajtu XORP-a kao minimalna zahtevnost (*dependence*) za kompajliranje je napomenut GMAKE. Skoro sve Linux distribucije već dolaze sa instaliranim GNU *make* programom, tako da često nije potrebno dodatno instalirati. Međutim, BSD ne integriše GMAKE u standardnu instalaciju, pa je neophodno prvo instalirati *gmake* paket. Na OpenBSD sistemu potrebno je izvršiti sledeće dve komande:

*export PKG_PATH=ftp://ftp.openbsd.org/pub/OpenBSD/4.4/packages/i386/ pkg_add gmake*

Skripta *configure* uspešno završava izvršavanje kada poslednjih nekoliko linija sadrži reči *Creating Makefile* ...

    Kada *configure* uspešno završi proveru sistema od svih neophodnih zavisnosti, može se preći na proceduru kompajliranja. Za razliku od drugih aplikacija u Linux-u, XORP kroz jednu komandu pokreće proceduru i za kompajliranje i za instalaciju nakon uspešnog kompajliranja. Procedura kompajliranja i instalacije pokreće se preko komande:

*gmake* ili *make*

Vreme kompajliranja zavisi od raspoloživih resursa. Na Alix ploči, kompajliranje je znatno brže nego na Soekris. Proces kompajliranja opterećuje procesor skoro 100% celo vreme kompajliranja. Zbog toga je potrebno obezbediti i dodatno hlađenje za *embedded* platforme. Često je dovoljan i mali hladnjak prislonjen na procesor. Poželjno je koristiti termalnu pastu za bolje provođenje toplote na hladnjak.

### VIII. POKRETANJE XORP PLATFORME

Nakon uspešnog kompajliranja automatski se pokreće instalacija kompajliranih datoteka u operativni sistem. Nakon toga, korišćenje USB fleš memorije gde je smešten izvorni kod više neće biti potrebna, jer su svi potrebni fajlovi na CF kartici.

    XORP platforma se sastoji od dva odvojena servisa. Prvi servis je glavni

servis koji vrši kompletnu funkcionalnost XORP softverskog rutera. Drugi servis je konzola za pristup konfiguraciji i kontroli glavnog XORP servisa.

XORP platforma se ne pokreće automatski po podizanju sistema, već je neophodno startovati glavni servis komandom:

*/usr/local/xorp/bin/xorp_rtrmgr*

Glavni servis traži konfiguracionu datoteku gde se nalaze instrukcije šta ruter treba da radi. Zbog toga je neophodno prvo napraviti konfiguracionu datoteku sa minimum konfiguracionih parametara, pre nego što se aktivira servis.

Potrebno je kreirati konfiguracionu datoteku preko komande:

*vi /usr/local/xorp/config.boot*

Zatim kliknuti na dugme INSERT i ubaciti sledeću minimalnu konfiguraciju:

```
interfaces {
    interface dc0 {
         description: "data interface"
         disable: false
         /* default-system-config */
         vif dc0 {
             disable: false
             address 10.10.10.10 {
                   prefix-length: 24
                   broadcast: 10.10.10.255
                   disable: false
             }
         }
    }
}
```

Umesto logičkog interfejsa *dc0* treba da stoji logički interfejs koji detektuje sistem npr. *eth0* ili *sis0*. Ovom konfiguracijom je ruteru saopšteno da za taj zadati interfejs koristi zadate mrežne parametre. Ukoliko se želi da interfejs koristi mrežne parametre koje mu je dodelio operativni sistem potrebno je da uklonite komentare sa *default-system-config* i iskomentarišete ceo *vif* blok.

Sa Shift+ZZ izlazite iz *vi* editora i snimate konfiguraciju. Treba obratiti pažnju, da ako je u operativnom sistemu podešena IP adresa, ista adresa bude podešena i u konfiguracionoj datoteci, u suprotnom mogu se desiti nepredvidivi problemi.

Potrebno je još dodati korisnika koji će XORP koristiti u operativnom sistemu.
Kod Linux-a se to radi komandom *useradd xorp*.
Kod OpenBSD-a preko komande *adduser xorp*.

Kada se ima ova minimalna konfiguracija može se podići glavni servis XORP-a komandom:

*/usr/local/xorp/bin/xorp_rtrmgr –b /usr/local/xorp/config.boot*

Videće se poruke kojima XORP obaveštava koje podsisteme je podigao u okviru softverskog rutera. Ukoliko se dobije linija koja sadrži *Error*, verovatno nešto u konfiguracionoj datoteci nije u redu.

Kada se servis podigne, i podigne sve potrebne module, softverski ruter XORP je podignut.

Konfiguracija rutera može se obaviti ili preko konfiguracionog datoteke ili iz servisa konzole. Lakši je pristup konfiguraciji iz konzole. Da bi se koristila konzola, otvori se nova SSH sesija ka uređaju i pokrene se konzolni servis:

*/usr/local/xorp/bin/xorpsh*

Ukoliko je glavni servis XORP-a startovan, konzola će se otvoriti i ponuditi svoj *shell*. Kretanje i mogućnosti konzole su slične kao i kod mnogih drugih rutera. Sa znakom *?* mogu se videti ponuđene opcije u aktuelnom modu, a sa dugmetom TAB izvršiti *autocomplete*.

Ako se želi da se izmeni konfiguracija rutera iz konzole, potrebno je uneti komandu *config* TAB ENTER, kako biste ušli u konfiguracioni mod rutera.

Ukoliko se dobije poruka o grešci, da se nema dovoljno prava da se izvrši konfiguracija potrebno je uraditi sledeće:
- Vratiti se u terminal u kom je startovan XORP glavni servis, i prekinuti proces sa Ctrl+C
- Editovati *gorup* datoteku tako da prihvata korisnika *root* u grupu *xorp*
  *vi /etc/group*
  pritisnuti taster INSERT za ubacivanje karaktera
  pronaći liniju *xorp:xxx:* i prepraviti na *xorp:xxx:root*
- Sačuvati datoteku sa SHIFT+ZZ
- Pokrenuti ponovo XORP servis
  */usr/local/xorp/bin/xorp_rtrmgr –b /usr/local/xorp/config.boot*
- U drugom terminalu pokrenuti konzolu
  */usr/local/xorp/bin/xorpsh*
- Pristupiti konfiguracionom modu

### IX. PODIZANJE XORP-A SA OPERATIVNIM SISTEMOM

Procedura podešavanja podizanja XORP-a sa operativnim sistemom zavisi od samog sistema. Generalno trebalo bi da radi sledeći postupak:

- Editovati fajl */etc/rc.local*
  *vi /etc/rc.local*
- Pre komande *exit* (ako postoji) dodati:

*/usr/local/xorp/bin/xorp_rtrmgr -b /usr/local/xorp/config.boot >> /var/log/xorp 2>&1 &*

- Upamtiti konfiguraciju za SHIFT+ZZ

Linija koja je upisana u startnu datoteku pokreće servis XORP sa zadatom konfiguracionom datotekom, ali i sve poruke koje dobija od XORP-a upisuje u *log* datoteku na lokaciji */var/log/xorp*. To je korisno jer se može pročitati *log* kada je potrebno komandom:

*tail /var/log/xorp*

Nakon *restart*-a uređaja, trebalo bi da se XORP podigne automatski. Pregledom *log* datoteke može se saznati da li je nešto krenulo onako kako nije trebalo.

### X. SIMULACIJA – MULTIKAST RUTIRANJE SA DVA RUTERA

Ova demonstracija prikazuje funkcionalnost XORP platforme kao *multikast* rutera u realnom okruženju sa dva rutera.

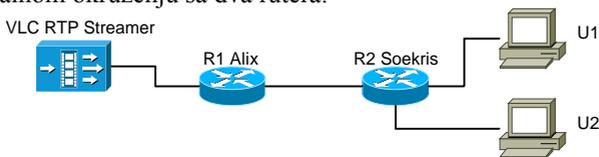

*Slika XI.1 Skica simulacije*

Cilj simulacije jeste da se omogući pregled *multikast* multimedijalnog toka podataka na računarima U1 i U2 koji se nalaze u udaljenim mrežama (Slika XI.1). Simulacija će pokazati funkcionalnost XORP platforme u vidu, *unicast* rutiranja, *multikast* rutiranja, IGMP protokol, PIM-SM protokol.

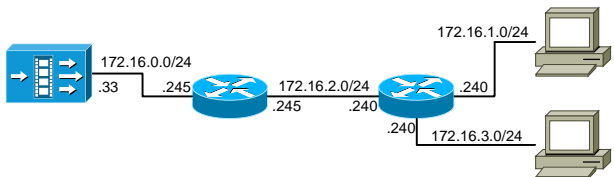

*Slika XI.2 Mrežna organizacija simulacije*

IP adrese U1 i U2 računara nisu bitne, bitno je samo da budu u zadatoj mreži i da im *default* ruter bude .240 . Na fizičkom nivou R1 Alix je sa eth0 portom povezan na *streaming* server, a eth1 portom na R2 Soekris. R2 Soekris je sa sis0 portom ka U1 računaru, sa sis1 portom ka R1 Alix ruteru, i sa sis2 portom ka U2 računaru.

Nakon fizičkog povezivanja uređaja prvi korak je obezbeđivanje RTP *multikast streaming*-a. Za tu potrebu koristi se aplikacija VideoLAN (VLC) koja je podešena da šalje film sa DVD-a na mrežu preko RTP *streaming*-a na *multikast* adresi 224.224.224.224. Video i zvuk se preko TS *mux*-era pakuju u isti paket. Paketi se šalju na port 1234.

Sledeći korak je konfiguracija rutera R1 i R2 za traženu funkcionalnost. Na ruteru R1 Alix prvo je potrebno konfigurisati IP adrese. Bolje je konfigurisati IP adrese u samom sistemu, jer u tom slučaju i ako se ne podigne XORP sistem će se odazivati na zadat IP. U sistemu Voyage Linux statičke IP adrese se definišu u datoteci */etc/network/interfaces.*

Nakon *restart*-a sistema ili samo mrežnog servisa, R1 Alix će uzeti zadate adrese. XORP konfiguracija za R1 može se pogledati na adresi: http://a.paxy.in.rs/radovi/AlixXORP.conf.

Da bi se obezbedila osnovna *unicast* povezanost podešena je statička rute na R1 ruteru ka mrežama .1.0 i .3.0 . Podignut je IGMP verzija 3 kao i drugi moduli koji su preduslov za korišćenje PIM-SM protokola. PIM-SM protokol predviđa korišćenje *randezvous point* (RP) rutera koji predstavlja tačku odlučivanja o članstvu nekog interfejsa određenoj *multikast* grupi.

PIM-SM poseduje protokol *bootstrap* koji se može koristiti za automatski izbor najbolje tačke za RP. Međutim, ovde se koristi jednostavna mrežna konfiguracija sa dva rutera, pa je jasno da je najbolja pozicija za RP upravo R1 Alix. Zbog toga je definisan statički RP za sve *multikast* adrese da ukazuju na ruter R1 Alix.

Na ruteru R2 Soekris takođe je prvo neophodno konfigurisati statičke IP adrese.To se radi kreiranjem datoteke za svaki interfejs. Npr za sis0 kreira se fajl */etc/hostname.sis0* i doda mu se linija sa IP adresom:

*inet 172.16.1.240 255.255.255.0 NONE*

Nakon *restart*-a sistema, interfejsi će uzeti konfigurisane IP adrese.XORP konfiguraciju za R2 može se pogledati na adresi: http://a.paxy.in.rs/radovi/SoekrisXORP.conf.

Dodata je statička ruta ka mreži .0.0 kao i RP za sve *multikast* adrese ka 172.16.2.245.

Sa klijentske strane potrebno je obezbediti osnovnu *unicast* povezanost sa *streaming* serverom podešavajući IP adresu, *subnet* masku, i *default gateway*.

Za U1 npr: 172.16.1.1 255.255.255.0 172.16.1.240
Za U2 npr: 172.16.3.1 255.255.255.0 172.16.3.240

Kako bi prihvatio i prikazao *streaming* klijent mora da poseduje aplikaciju za *multikast streaming*. Korišćen je VLC kako za slanje, tako i za prijem *multikast streaming* video paketa. U VLC klijentu potrebno je povezati se na url:

*rtp://172.16.0.33@224.224.224.224:1234*

VLC saopštava preko IGMP protokola da želi da prima *streaming* na *multikast* adresi 224.224.224.224 gde je izvor 172.16.0.33 koji je RTP tipa na portu 1234.

Simulacija je uspešna kada se na klijentskoj strani dobije video i zvuk. Isto se može uraditi i za U2.

Treba primetiti, pre pokušaja povezivanja, preko LED indikatora na Ethernet portu vidi se da postoji masovan saobraćaj od *Streaming* servera do R1, ali ne i ka R2 i U1 i U2. Tek nakon povezivanja U1 vidimo indikaciju da masovan saobraćaj postoji između *Streaming* servera, R1, R2 i U1, ali ne i U2. Nakon priključivanja i U2, dobijamo kompletan *multikast* saobraćaj na svim interfejsima.

## XI. ZAKLJUČAK

*Multikast streaming* prestavlja ne iskorišćen potencijal u računarskim mrežama. Njegovo je svojstvo da slanjem jednog primerka paketa usluži neograničen broj korisnika doprinosi. Njegovo korišćenje tek polako postoje aktuelno kroz 3Play mreže, gde se distribucija radio i TV signala vrši isključivo članstvom u *multikast* grupama.

XORP platforma je definitivno jedan od uzora implementacije *multikast* rutiranja. Neki od najpopularnijih softverskih rutera (kao na primer MikroTik) baziraju *multikast* funkcionalnost upravo na ovoj platformi.

XORP platforma poseduje vrlo kvalitetnu projektnu dokumentaciju, kao i API za dalji razvoj. Uz određeno prilagođavanje, može se koristiti za implementaciju hibridnih hardversko-softverskih rutera.

## XII. KOMENTAR

U srpskom jeziku se koristi veliki broj stranih reči. Oblast računarskih mreža obuhvata mnoštvo reči koje bi prevođenjem izgubile originalno značenje.

Zbog toga određeni termini su zadržani u izvornom obliku, a njihovo objašenjenje je dato na kraju rada.

### XIII. POJMOVI

[1] Multikast (višestruko upućivanje) – predstavlja način prosleđivanja jednog paketa na više odredišta koristeći jednu odredišnu adresu. Pošaljilac šalje jedan primerak paketa na multikast adresu. Svi koji žele da prime taj paket prihvataju paket poslat na tu adresu.

Primer: Pošta je poslata na adresu Karađorđeva 13, dostavljena je do vrata zgrade, ali nije adresirana ni na koga posebno. Svaki stanar te zgrade moze da uzme tu poštu ako želi.

[2] Multikast rutiranje – predstavlja metodu kojom ruter uči putanju kojom treba proslediti paket poslat na odgovarajuću multikast adresu. Ovime se obezbeđuje da primerak paketa bude dostupan klijentima kada požele da a prime.

[3] Embedded platforma – hardverska podloga sa integrisanim komponentama kao što su procesor, RAM memorija, fleš memorija, mrežni i USB interfejsi, itd. To su uređaji koji predstavljaju osiromašenu varijantu PC računara. Cilj ovih uređaja je da izvršavaju operacije aplikacija.

### LITERATURA

### ABSTRACT

With integration of software routers with embedded platform we can get the most modern routers, but with less price. TV and radio services over IP networks become available just when multicast routing is used. Multicast routing is currently feature only on expensive hardware solutions. XORP open-source platform offers multicast routing through a software router, with the possibility of integration of cheap embedded systems.

**Mutlicast routing open-source platform – XORP**
Petar Bojović, Katarina Savić, Aleksandra Smiljanić